# Opinion Crystallography: Polarizations, Symmetries, Bonds, and Bands

Çağlar Tuncay
Department of Physics, Middle East Technical University
06531 Ankara, Turkey
caglart@metu.edu.tr

**Abstract:** May randomness (real numbers, opinions) evolve into order (regularity) with time? We study some polarization and symmetry properties, which emerge in time evolution of opinions (real numbers) within entries of two and three-dimensional lattices, which had initial randomness.

**1 Introduction:** Entries (i) of a lattice (M=NxN for square, and M=NxNxN for cubic lattices) have random (homogeneous) distribution of initial opinions ($Q_i(0)$) in terms of real numbers. In usual iterative interaction tours (time, t) only the nearest neighbor (nn) entries interact and opinions ($Q_i$) average [**1**];

$$Q_i(t) = (Q_i(t-1) + \sum_j^{nn} Q_j(t-1) )/(\rho+1) \quad , \qquad (1)$$

where ρ is the number of nn for (i). We follow parallel (synchronous) updating.

The average opinion for the lattice ($Q_{av}(t)$) at a given interaction tour be given as

$$Q_{av}(t) = Q^T(t)/M = (\sum^M_i Q_i(t) )/M \quad , \qquad (2)$$

where $Q^T(t)$ is the total opinion at a given time, i.e., $Q^T(t) = \sum^M_i Q_i(t)$, and $Q_{av}(t \rightarrow \infty)= Q_{av}(\infty)$.

Are $Q_i(t)$ (Eqn. (1)) random for $0<t<\infty$, with random $Q_i(t=0)$? Or, may $Q_i(t)$ involve some polarization and symmetry for $0<t<\infty$?

In a recent paper [**2**], we had observed some unexpected polarization and symmetry features in lattices during time evolution of opinions. In this contribution we study them in detail.

**2 Order (regularity), polarization and symmetry:** We announce that, initial randomness in opinions is observed to evolve into order (regularity) with time, under certain conditions. Performing parallel (synchronous) updating (Eqn (1)), we observed the initial random opinions evolving into order (regularity). Opinions are observed to display polarization for $t<\infty$; and what we mean by *polarization* is the order (regularity) within opinions as follows: One corner or edge of the two-dimensional lattice involves opinions greater than $Q_{av}(t)$, and the opposite corner or edge involves these less than $Q_{av}(t)$, or vice versa. One may define $Q'_i(t)$ as, $Q'_i(t)=Q_i(t) - Q_{av}(t)$, and the mentioned *symmetry* (and polarization) properties come out about zero (since, $Q_{av}(t)\rightarrow Q_{av}(\infty)$, as $t\rightarrow\infty$) for $Q'_i(t)$. Moreover, distribution of opinions at $t<\infty$ over the entries are observed to be invariant (up to numerical approximations performed by the software that we utilized) under many symmetry operations such as rotation of the lattice by (some integer multiples of) $\pm\pi$ about body axis and about bisector of any edge, inversion, etc. And, accuracy (up to the mentioned approximations) within such symmetries are observed to increase with time, as $Q_i(t)\rightarrow Q_{av}(\infty)$, for $t\rightarrow\infty$ and $Q_{av}(t)\rightarrow Q_{av}(\infty)$, for $t\rightarrow\infty$; similarly, as $Q'_i(t)\rightarrow Q'_{av}(t)$ for $t\rightarrow\infty$ and $Q'_{av}(t)\rightarrow Q'_{av}(\infty)=0$ for $t\rightarrow\infty$. All of the polarization and other symmetry features decay ultimately.

Figure 1 is the evolution of $Q_i(t)$ (Fig 1a), and that of $Q_{av}(t)$ (Fig. 1a), for a 8x8 lattice; where, $Q_{av}(t=200)=0.0019=\sim Q_{av}(t\rightarrow\infty)$. Please note that, $Q_{av}(t)$ varies (either decreases or increases (Figs. 3 and 4) linearly in logarithm of time, (as arrows in the relevant figures designate) for the first ten tours (t) or so. In many runs we observed that the slope of the mentioned variation in $Q_{av}(t)$ decreased with increasing N, i.e., we observed steeper decreases or increases for smaller N, where minor modulations may appear due to initial randomness. Figure 2 displays the polarization within the same lattice at t=150, where parallel (synchronous) updating is performed in terms of Eqn. (1).

We investigate polarization and symmetry properties of opinions in two parts: Firstly we study two-dimensional lattices. Later, we consider three-dimensional lattices, in terms of faces and bulk.

**3 Applications and Results:** Initially we charge lattices homogeneously with random real numbers ($Q_i(0)$, where $-1.0\leq Q_i(0)<1.0$). And, let the opinions evolve in terms of averaging process [**1**] (Eqn. (1)), where we apply parallel (synchronous) updating.

## 3 a Two dimensional lattices

In this section we consider 10x10 and 9x9 lattices, with even and odd dimension, respectively.

For a square lattice, ρ=4 (in Eqn. (1)) if (i) is in the square, ρ=3 if (i) is on an edge, and ρ=2 if (i) is at a corner.

In Figures 3, and 4 we show time evolution of $Q_{av}(t)$ (Eqn. (2)), for 10x10 and 9x9 lattice, respectively. Please note that $Q_i(t)$ varies linearly in logarithm of time in all; $Q_i(t)\rightarrow Q_{av}(\infty)$, as $t\rightarrow\infty$. Similarly $Q_{av}(t)\rightarrow Q_{av}(\infty)$, as $t\rightarrow\infty$. (Also see, Fig. 1 b.) We observe that, convergence of $Q_{av}(t)$ is slower for bigger dimension (N). Secondly, lattice sum of the opinions ($Q^T(t)=M\ Q_{av}(t)$, Eqn (2)) is not conserved through the evolution, i.e., $Q^T(t\neq0)\neq Q^T(t=0)$; also in Fig. 1 b.

**Observations:**

1) $Q_i(t)\rightarrow Q_{av}(\infty)$ as $t\rightarrow\infty$, (i.e., $Q'_i(t)\rightarrow0$ as $t\rightarrow\infty$) independently of the sub index and the number of dimension of the lattice (N). (Figs. 1, 3, and 4.)
2) Initial randomness within opinions does not survive, and it decays rapidly, within about ten tours for N≤10 (Figs. 1, 3, and 4.). We name this period of time the short run (first epoch). Afterwards, patterns (variation of opinions along any edge from entry to entry) start to display regularities, and we observe repeated patterns. Whenever the polarization in lattice develops, one end of an edge of the lattice may involve a greater opinion than the other end (or, vice versa). We name this period of time the intermediate run (second epoch). As time evolves the patterns smooth and $Q_i(t\rightarrow\infty)\rightarrow Q_{av}(\infty)$; we name this period of time as long run (third and ultimate epoch).
3) For intermediate run, we have clear polarization within $Q_i(t)$. Namely, $Q_i(t)$ varies smoothly and continuously (about $Q_{av}(t)$ in the second epoch, and about $Q_{av}(\infty)$ in the third epoch) from one corner of a square of the lattice to diagonally opposite one; from a higher value to a lower one, or vice versa. Polarization may occur also from one edge of a square to the opposite edge (or at least approximately), where one may rotate the opinions on any pair of opposing edges by π, about the corresponding Cartesian axis (or perform other symmetry operations) and obtain the opinions on the opposite edge. We performed many runs for many t and N; and observed the mentioned case in many of them, as expected. (See, Figs. 3. Others are not shown.)
4) We run our evolutions many times, with different random initials $Q_i(t=0)$, i.e., for various a, and b in $a\leq Q_i(0)<b$, and observed similar results as mentioned in previous cases. Sum and difference of opinions along edges too came out as independent of lower and upper bounds for initial random real numbers; yet, with different patterns. Some other linear (and possibly non-linear) combinations of opinions may display several types of (other) symmetries.

5) We had randomness at the beginning, $Q_i(t=0)$; and we have ultimate equality, $Q_i(t\rightarrow\infty)=Q_{av}(\infty)$, and probability density function (PDF) for $Q_i(0)$ was (almost) constant (due to homogeneity in random numbers). We have delta PDF ultimately (not shown).

We may explain the situation as: Origin of the mentioned symmetries in opinions is the symmetry properties of the lattice under consideration and symmetries are induced onto opinions by means of averaging process given (Eqn. (1)), where the given expression is symmetric with respect to (i), and nn entries. And, discontinuity at the borders of the lattice may be neglected for large N. We have perfect cubic symmetry properties ultimately, i.e., as $t\rightarrow\infty$. Yet, as $t\rightarrow\infty$, $Q_i(t)\rightarrow Q_{av}(\infty)$.

## 3 b Three dimensional lattices

In this section we consider cubic lattices with various (N) (figures are not shown).

For a cubic lattice, $\rho=6$ (in Eqn. (1)) if (i) is in bulk, $\rho=5$ if (i) is on a cubic face, $\rho=4$ if (i) is on a cubic edge, and $\rho=3$ if (i) is at a corner. And, we follow parallel (synchronous) updating.

Within all the figures we obtained for $Q_i(t)$ varying linearly in logarithm of time; $Q_i(t)\rightarrow Q_{av}(t)$, as $t\rightarrow\infty$. Similarly $Q_{av}(t)\rightarrow Q_{av}(\infty)$, as $t\rightarrow\infty$. And, the convergence is much slower, when compared with the case for two-dimensional lattices with the same N.

As in the two-dimensional lattices, we have three evolution epochs present, and in the second epoch (intermediate run) we observe similar polarization and symmetry relations within cubic lattices as we observed in square ones. Polarization emerges on each (square) face and within bulk; so, it continues from one face to the opposite face through the bulk. In the same manner, we have polarization along any line connecting the surfaces. In the third epoch symmetries and polarizations all decay with $Q_i(t\rightarrow\infty)\rightarrow Q_{av}(\infty)$.

In summary, we have polarization and symmetry within opinions in three-dimensional lattices, which are very similar to the case in two-dimensional lattices; yet, we have more symmetry relations (due to the point group of cubes, which is richer than that of squares).

**3 c Extraordinary cases:** We tried three extraordinary cases, which are:

1) Extra ordinary initiation: We fed the lattice with random numbers only for some (i), and left the other entries as zero. We confined the nonzero portion about the center, as a square with dimension $n<N$, for two-dimensional lattices, and as a cube with dimension $n<N$, for three-dimensional lattices. We also performed many runs, where the nonzero portion was about any corner. And we obtained polarization and symmetry in all of these cases. Furthermore we tried some functions as $Q_i(0)=\sin(I/N) + \sin(J/N)$, and $Q_i(0)=(I/N) + (J/N)$, with (I,J)=(i) for two-dimensional lattices, and observed polarizations and symmetries in all. One may expect similar results for three-dimensional lattices.

2) Random interaction: We selected the averaging entries randomly (instead of following Eqn (1)); where, at each tour, the opinions involved by only two randomly selected entries ((i) and (i'), say) are averaged. In this case, we did not observe any polarization and symmetry features within the evolving opinions.

3) Broken bonds: It may be underlined that, only the nn sites are considered up to now, and we observed (two- and) three-dimensional (square) cubic symmetries within opinions. We had only nn connections, with equal bond strengths. We run for several cases, where we broke two neighbor bonds (i.e., we ignored the two neighbor bonds, and considered the other two neighbor bonds) and we broke two opposite bonds, and thirdly we broke three of the nn bonds, for various N in two- and three-dimensional lattices. And, we did not obtain polarization and symmetry when we had any three and two neighboring bonds are broken, i.e., we obtained polarization and symmetry whenever two opposite bonds are broken. Figure 5 displays results for (asymmetrically) broken two neighbor bonds (Fig. 5a), and (symmetrically) broken two opposite bonds (Fig. 5b), for N=10 at various t, as designated in the figure caption.

**4 Discussion and conclusion:** We performed many runs involving two-dimensional lattices with N=100 at maximum, and three-dimensional lattices with N=20 at maximum and detected similar observations as the ones mentioned within the previous sections.

We predict that these observations are independent of the distribution of initial random real numbers and their upper and lower bounds, yet the evolving patterns may be different. That is, the patterns all come out arbitrarily at different runs with fixed parameters (for number of tours and dimension) depending upon the randomness in $Q_i(t=0)$; we have polarization and symmetries emerging each time, which are independent of $Q_i(t=0)$.

Eqn. (2) is not conservative, i.e., $Q^T(t) \neq Q^T(0)$, so $Q_{av}(t) \neq Q_{av}(0) \neq Q_{av}$. In another set of runs, we utilized positive initial opinions ($0 \leq Q_i(0) < 1.0$), and obtained similar results.

We may define opinion as any (mathematical) quantity which evolves into regularity (with some symmetry properties) out of initial randomness. Within the present framework, we may consider the issue in the following way: Let's define a finite set $R(t=0)=\{r_i\}$, composed of random real numbers (opinions) $a \leq r_i < b$, with a, $r_i$, and $b \in R$, where R is the set of real numbers. If O is any algebraic operator composed of addition, subtraction, division and multiplication, the set S ($=O^n R$, and $S=R(t=n)$) which is obtained by n successive application of O on R may involve some symmetries. Please note that Eqn. (1) defines an algebraic operator, and $-1.0 \leq Q_i(t=0) < 1.0$, with random $Q_i(t=0) \in R$.

**FIGURES**

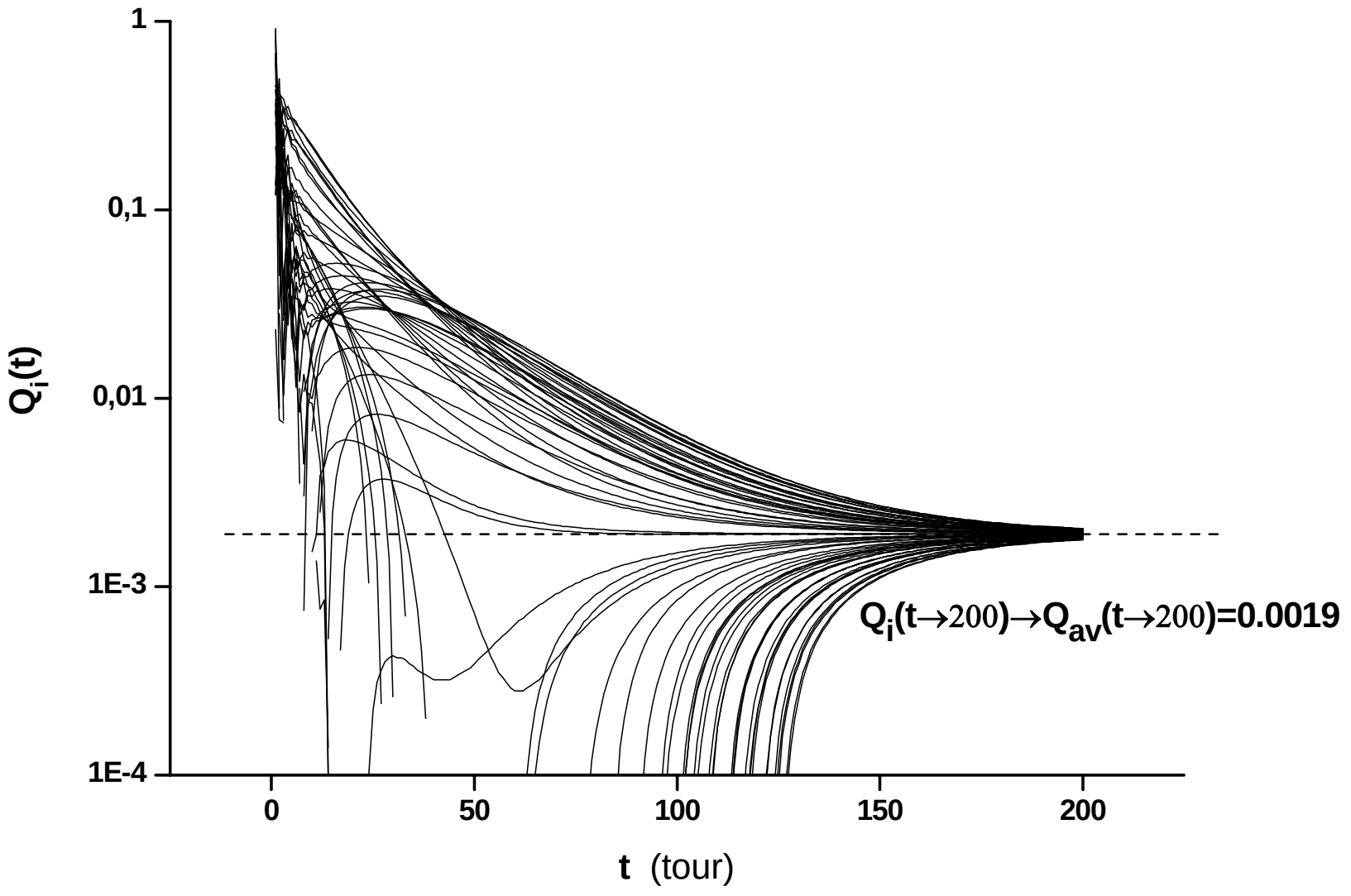

**Figure 1 a** Evolution of $Q_i(t)$ for a 8x8 lattice in 200 interaction tours, where the initial opinions are random.

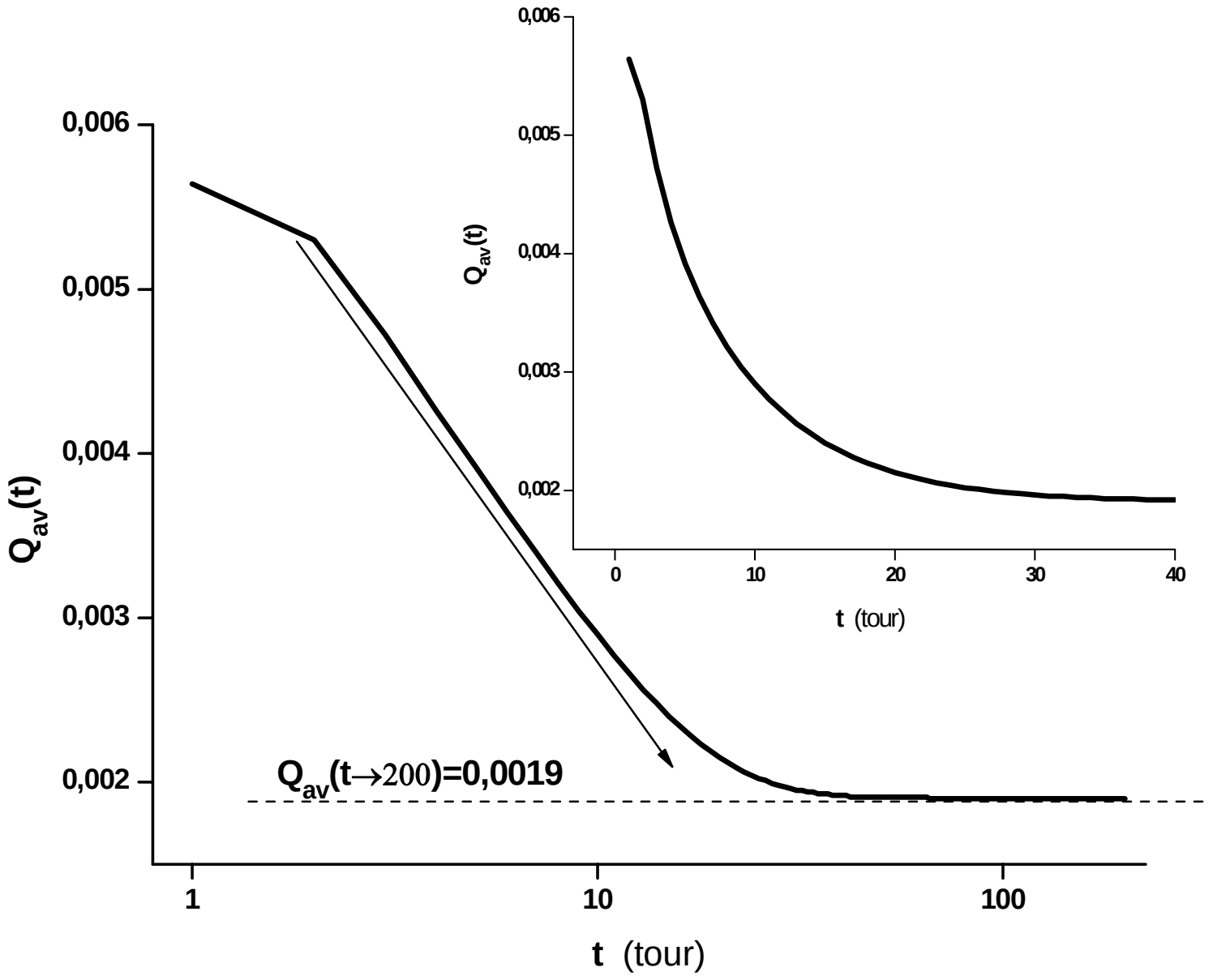

**Figure 1 b** Evolution of $Q_{av}(t)$, i.e. average value of opinions, parameters are as in Fig 1 b. The dashed line is for $Q_{av}(t=200)=0.0019$ in all, and inset in the $Q_{av}(t)$ plot shows the enlarged view for rapid convergence of $Q_{av}(t)$.

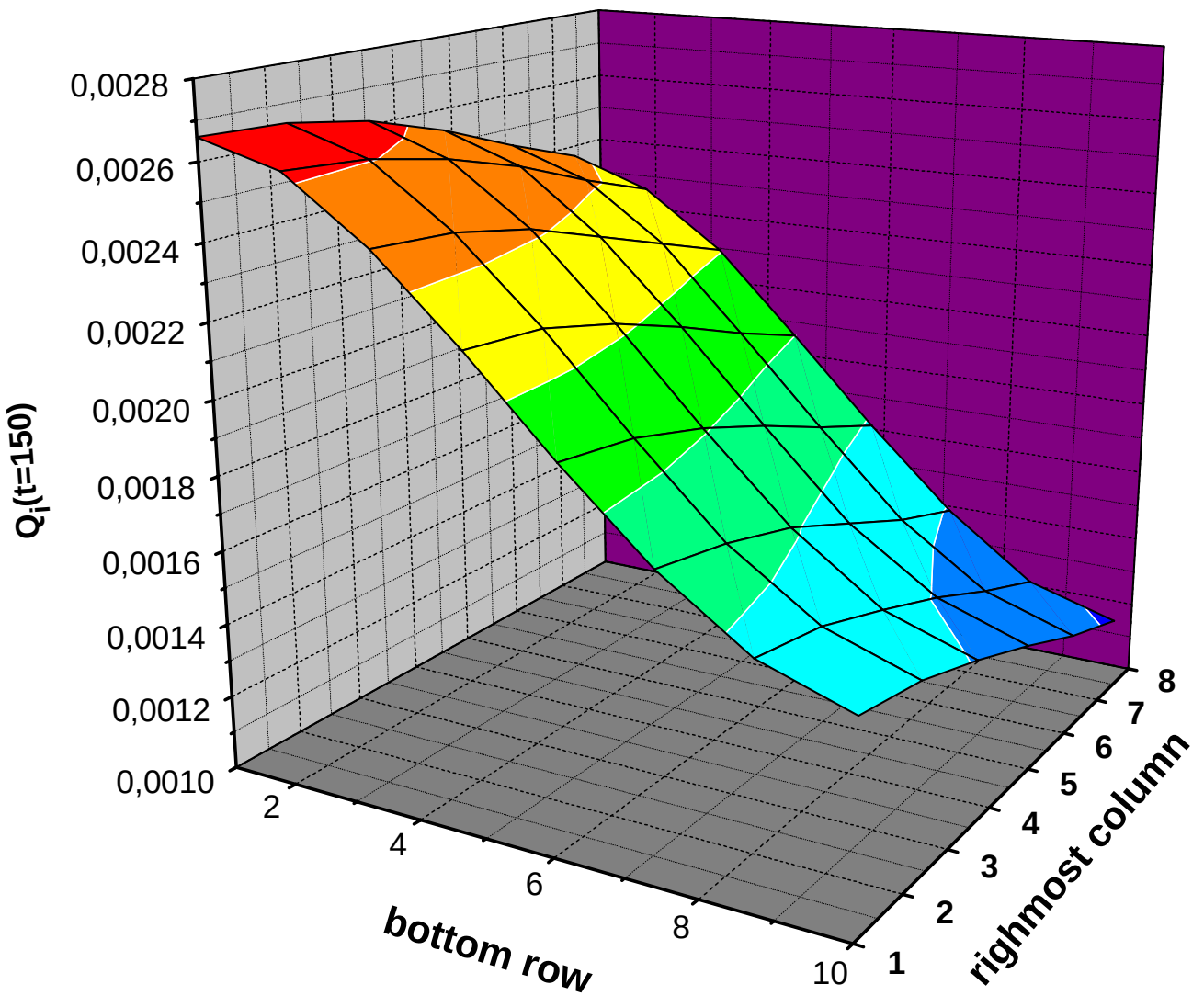

**Figure 2 a** Opinions of an 8x8 lattice in a three-dimensional contour plot at t=150. Vertical axis is for opinion, and the horizontal axes are for the edges of the lattice. Patches in color represent intervals of real numbers on the vertical axis. Patches in grey are for the view of the other (back) side (from below).

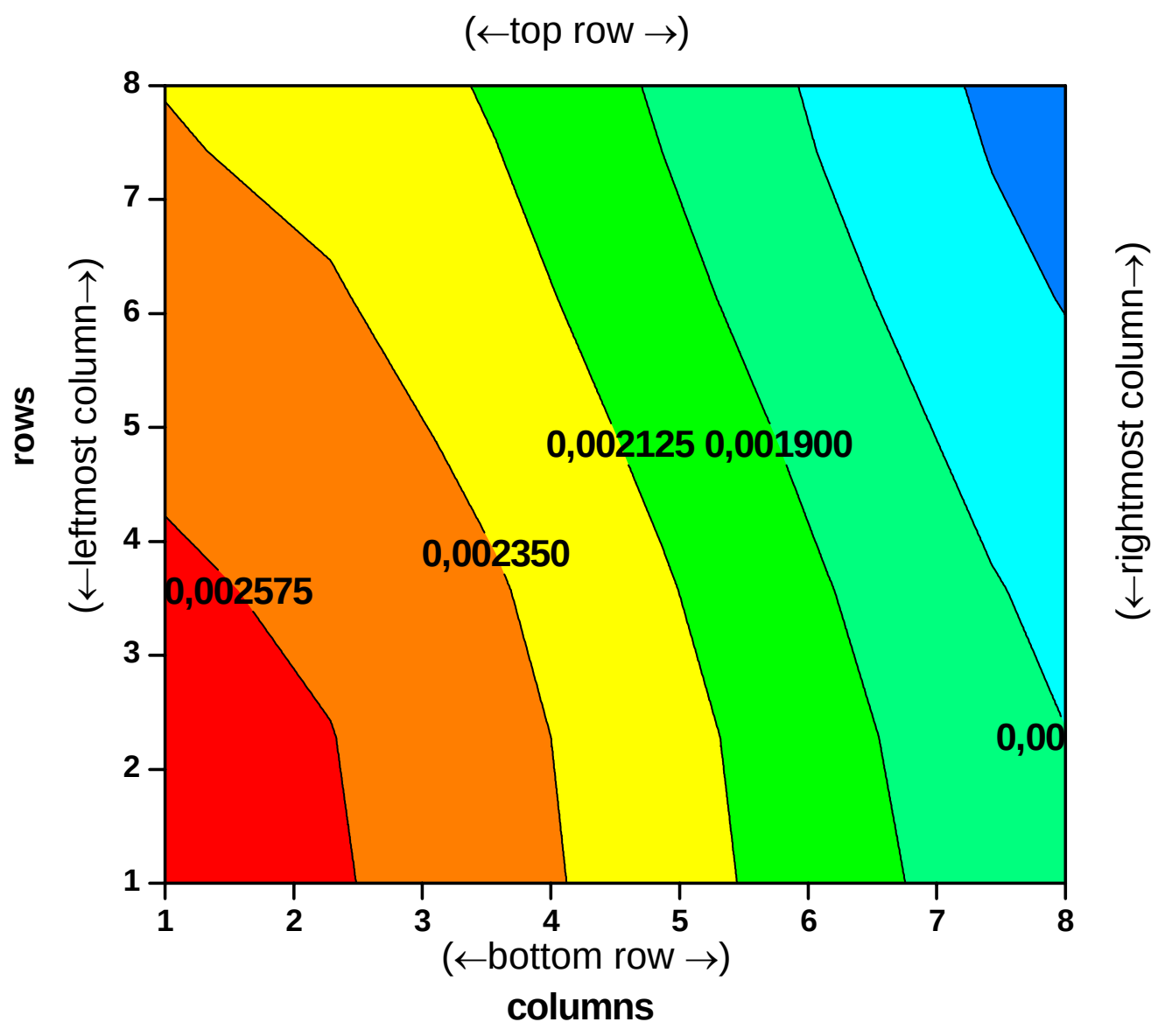

**Figure 2 b** Opinions on a 8x8 lattice in terms of two-dimensional contour lines at t=150, where the axes are for the edges of the 8x8 lattice, and opinions are designated in color patches with the values given on contour lines.

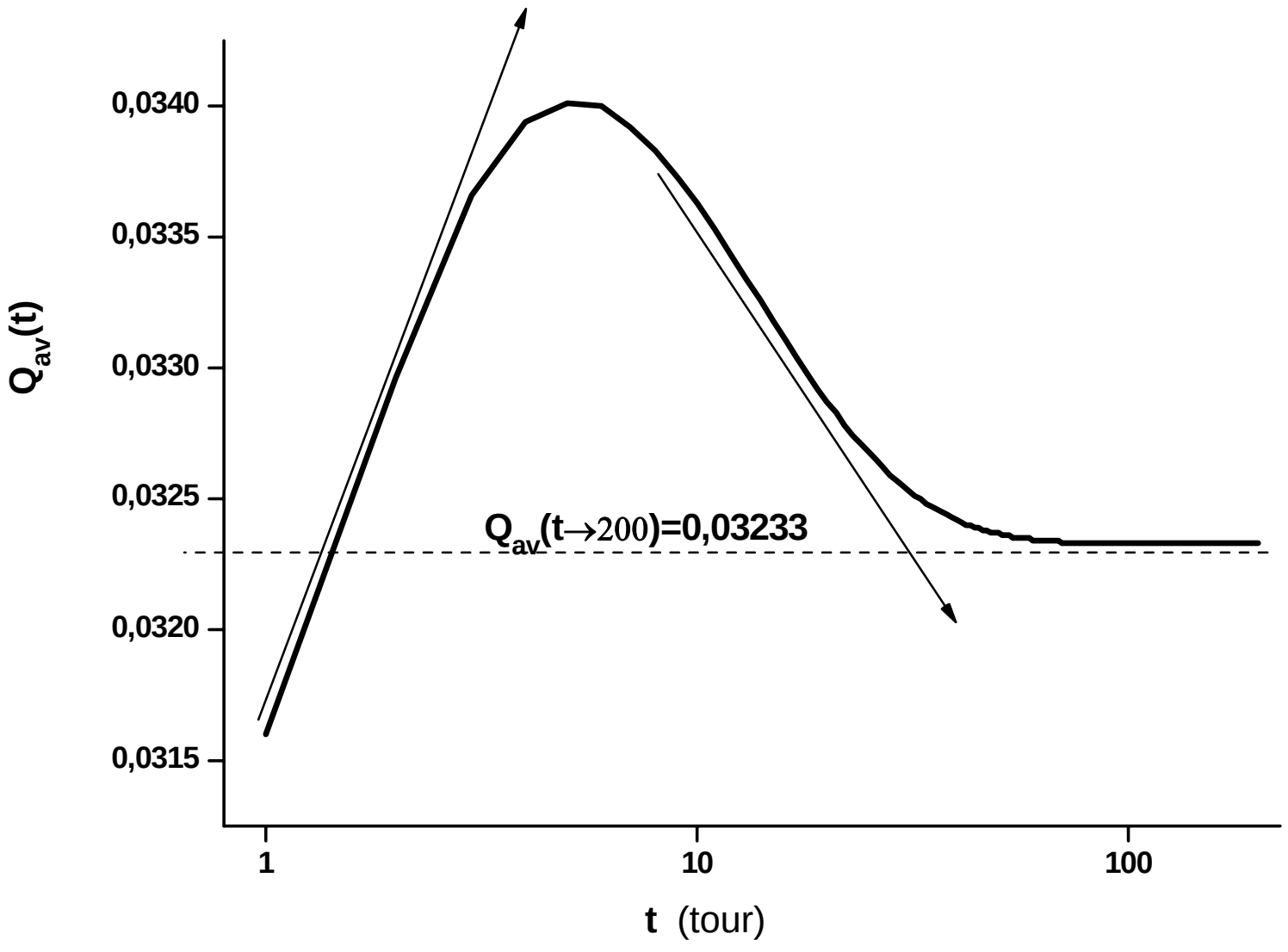

**Figure 3** Evolution of $Q_i(t)$, i.e. average of opinions for a 10x10 lattice in 200 interaction tours, where the initial opinions are random. The dashed line is for $Q_{av}(t=200)=0.03233$.

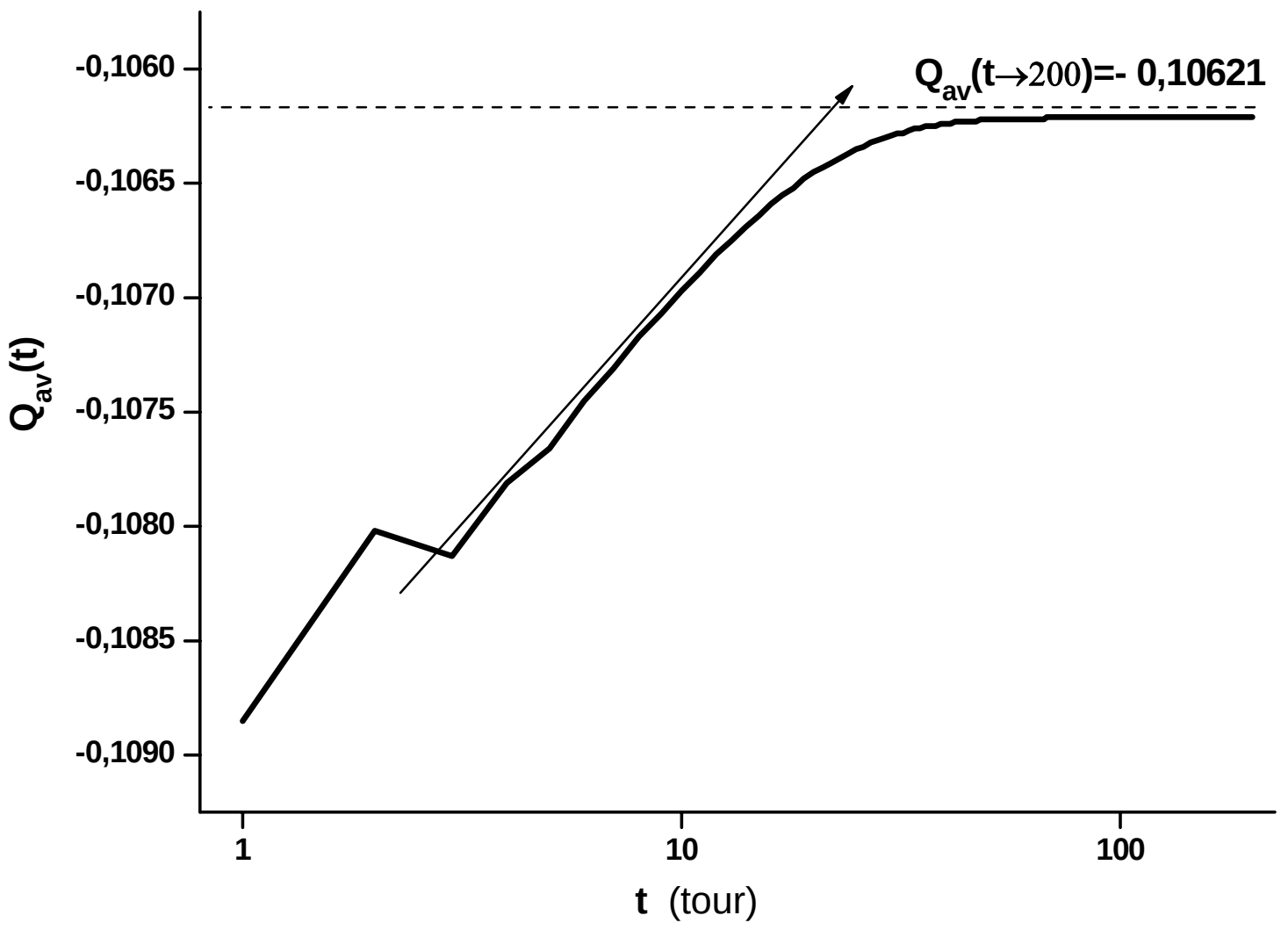

**Figure 4** Same as Figure 6, for a 9x9 lattice, where the dashed line is for $Q_{av}(t=200) = -0.10621$.

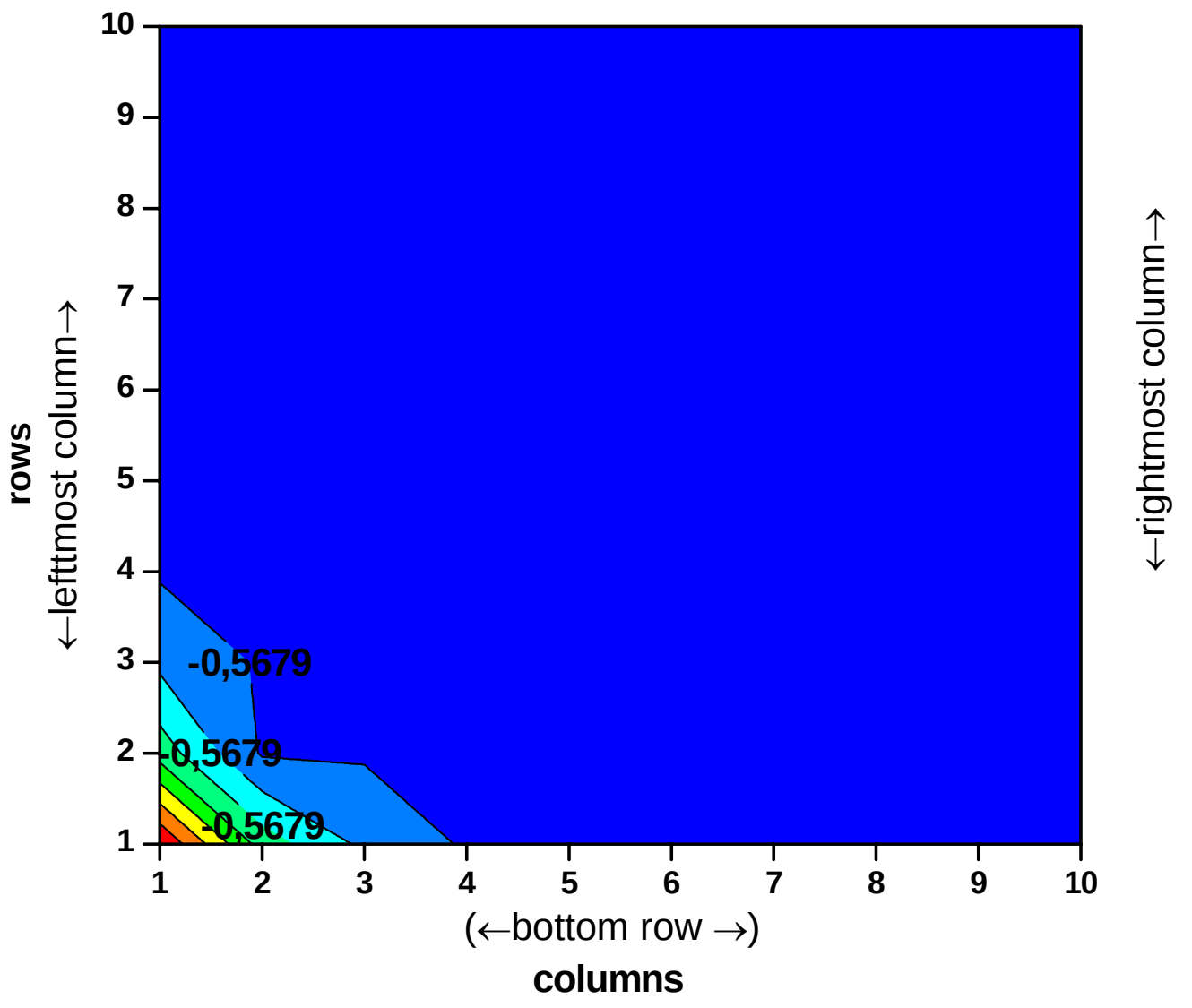

**Figure 5 a** Opinions in a 10x10 lattice at t=50, where two neighbor bonds are broken, and after a short period of time (few tours, t) we have equality in opinions everywhere within the lattice.

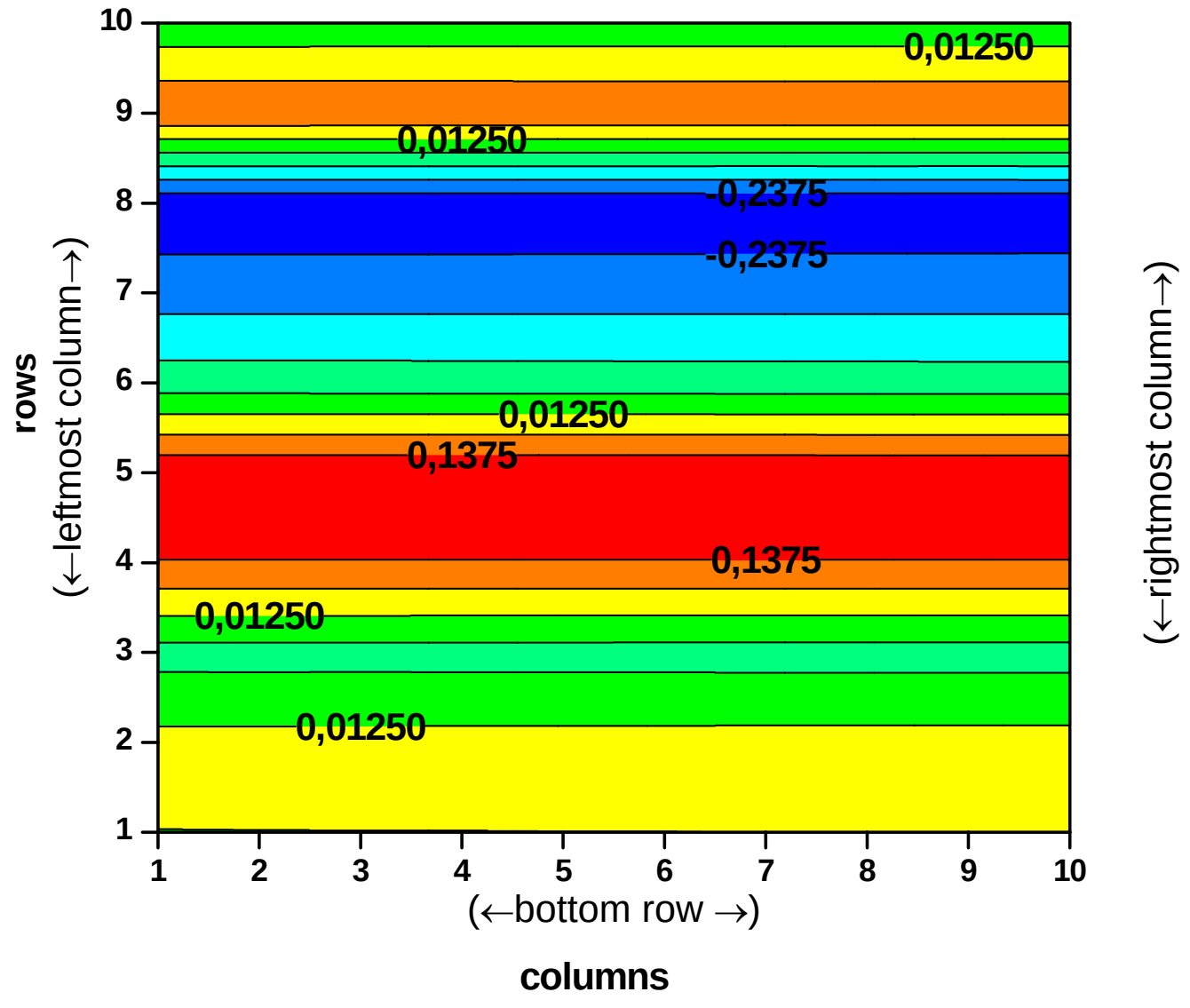

**Figure 5 b** Same as Fig 5 at t=150. Here, the two opposite bonds are broken.